# IN-NETWORK AGGREGATION USING EFFICIENT ROUTING TECHNIQUES FOR EVENT DRIVEN SENSOR NETWORK

Smitha N. Pai[1], K.C. Shet[2] and Mruthyunjaya H.S.[3]

[1]Dept. of CSE, M.I.T., Suratkal
[2]Comp. Engg., NITK, Suratkal
[3]Dept. Of E and C., M.I.T., Manipal University, Manipal

*ABSTRACT*

*Sensors used in applications such as agriculture, weather , etc., monitoring physical parameters like soil moisture, temperature, humidity, will have to sustain their battery power for long intervals of time. In order to accomplish this, parameter which assists in reducing the consumption of power from battery need to be attended to. One of the factors affecting the consumption of energy is transmit and receive power. This energy consumption can be reduced by avoiding unnecessary transmission and reception. Efficient routing techniques and incorporating aggregation whenever possible can save considerable amount of energy. Aggregation reduces repeated transmission of relative values and also reduces lot of computation at the base station. In this paper, the benefits of aggregation over direct transmission in saving the amount of energy consumed is discussed. Routing techniques which assist aggregation are incorporated. Aspects like transmission of average value of sensed data around an area of the network, minimum value in the whole of the network, triggering of event when there is low battery are assimilated.*

## 1. INTRODUCTION

Sensor devices are used to measure physical parameters like pressure, temperature, humidity etc. When placed within the transmission range of each other, it forms a sensor network. It carries the task of sensing, computation and forwarding. They have some limitations like computation, memory and energy. Sensors deployed in applications like the agricultural field require that the batteries be operating for one cropping season. Energy in the battery can be saved by reducing the number of transmissions and receptions. Reduction of the packet size, or distance between the nodes can also help in saving sufficient amount of energy. Efficient routing algorithms will have to be incorporated to find paths which consume minimal energy during path establishment and data transfer [1, 2].

The current paper is based on an on-going project COMMON_Sensewhere the water level is monitored using sensor network [3]. Sensors obtain energy to operate using solar energy, power grid lines or the battery. In applications like agriculture, the fields are away from the main land and so power grid lines are difficult to obtain. Solar panels when placed in the agricultural field will not only block the panels because of the large leaves, also there are chances of theft. Hence battery cells are the next available options.

Sensors deployed in the agricultural field are less in number and they are placed relatively far from each other. This reduces the cost factor on the farmers. They are basically used to monitor the water level in the field. Some sensors are also used to measure humidity. Information regarding the water level in the field if provided to the farmer could help them in regulating the flow of water. This avoids salinization of soil when the water is in excess. It can also provide only sufficient amount of water and hence improve on the yield of the crop. Over irrigating leads to disturbing the hydrology of the water bed which is undesirable.

## 2. RELATED WORK

Sensors deployed in the agricultural field read the soil moisture and report this data information to a common access point called the base station or the destination node. All the nodes in the network form the source node. One node among them which is approximately placed at the centre of the network is the base station. A path has to be established between all these nodes and the base station. In order to achieve this routing algorithms are essential. Nodes use these paths to send data continuously unless the path fails either due to atmospheric reasons, battery, electronic failure etc. Energy is expended in finding new path.

Usage of sensors in precision agriculture is common[4-7]. The COMMON-Sense project uses AODV protocol to establish the path between the source nodes up to the destination. In the current work better routing techniques are added so that path once established is active for long intervals of time. This avoids repeated path search and assist in saving energy.

In the field, few sensors are used and they are placed far away from each other. This can avoid expending more money on sensors. As the distance to the base station is too far off, the data has to be transmitted using the multi-hop technique.To avoid multiple transmissions of same data the whole network is partitioned into clusters, and data within a cluster is aggregated and sent to its cluster head.

Lot of work pertaining to routing, clustering and aggregation is carried out. A protocol based on the residual energy and minimum transmission cost (minimum hop count) along a path is projected [8]. Using minimum spanning tree a routing of the path is designed [9]. A genetic algorithm based routing protocol where a part of the route is swapped and new routes are established is addressed [10]. Anunequal clustering approach along with minimum hop inters cluster communication to achieve minimum total energy consumption in the network is suggested [11]. The network is partitioned into rings along with creating sectors within each ring. This sector is changed dynamically to form cluster with different set of nodes and send data using minimum hop [12].In the case of wheel based event triggered data aggregation, aggregation is done along the spokes of the wheel [13]. Aggregation along the levelled ring structure is carried out. Ring structure with aggregation along a ring is sent to the next layer of the ring is suggested [14]. A fine grained aggregation where wedges are created in the network and data is aggregated from the outer hop to the inner hop aggregating along the way to the base station [15]. Choice of route should be such that it takes lesstime to reach the base station along with aggregation taking place at the first level of the tree is projected. [16]. The aggregation and routing techniques studied help in designing a path to the base station, such that it lives for long, once established.

## 3. DEPLOYMENT

Broadcasting of control signals unnecessarily to large number of neighbours during path search can be avoided by choosing efficient topology for deployment. This is achieved by placing sensor such that it is heard either by three or four of its neighbours during transmission. This is the minimum number of nodes that can be placed close by so that the transmission is continued with at least three or four paths available all the time. This also ensures minimum usage of sensors, if the cost of the sensor network is expensive. These sensors form the *coordinator* nodes to send data to the neighbours using multi hop. Ref. [17, 18] relates to placement of sensor in grid topology. A comparison of triangular, square and hexagonal topology is carried out [17]. The triangular topology though a reliable network with longer life, requires large number of nodes. Square and hexagonal topologies require almost same number of nodes, with hexagonal requiring a little less number of nodes at the expense of reliability. Both have almost the same life span. Triangular topology sends data to the neighbouring six nodes, expending more energy than grid with four neighbours and hexagonal with three neighbouring nodes. In the current work square and hexagonal topology is used. Figure 1 shows the various topologies.

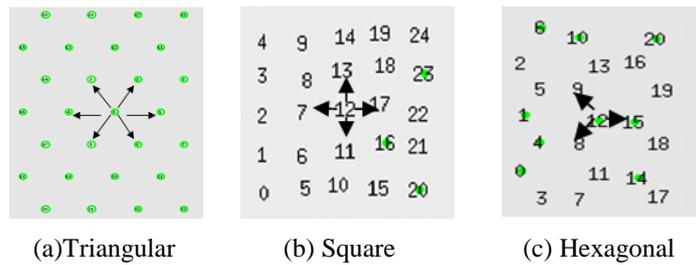

(a)Triangular    (b) Square    (c) Hexagonal

Figure. 1 Various topologies

## 4. SIMULATION PARAMETERS FOR AGRICULTURE

Parameters are designed using Tiny node. This mote is used in COMMON-Sense project. Table 1 shows the data sheet for Tiny node. The values marked in the table refers to the parameter values used in this study. Frequency of operation is 868Mhz, receiver sensitivity of -104dBm with transmit power of 5dBm, with antenna height of 1m and minimum transmission range of 200m. Current consumption during transmit, sleep and receive include 33mA, 1μA and 14mA. These values are again used for the computation as showed in the table 2.

Table 1 Tinynode data sheet [20]

| Parameter | Value |
|---|---|
| Operating Frequency | |
|     868 MHz version | 868-870 MHz |
|     915 MHz version | 902-928 MHz |
| RF Output Power | 0 to + 12 dBm |
| Data Rate | 1.2 – 152.3 kbps |
| Receiver Sensivity | |
|     @ 1.2 kbps | -121 dBm |
|     @ 76.8 kbps | -104 dBm |
|     @ 152.3kbps | -101 dBm |
| Range @ 76.8 kbps | |
|     Outdoor (1m elevation) | 200 m (+5 dBm) |
|     Indoor | 40 m (+5 dBm) |
| Current Consumption | |
|     Transmit @ +5 dBm | 33 mA |
|     Receive | 14 mA |
|     Sleep | < 1 uA |

## 5. ROUTING

Nodes are randomly deployed in the rectangular field. Efficient routing techniques are essential to enhance the life time of the sensor network. Some nodes are identified as coordinator nodes which approximately lie in the centre of a square or hexagonal region and act as aggregator node. These coordinator nodes are as planted as per the Figure 1. Packets are transferred between coordinator nodes using multihop. Other nodes in the network are associated with the coordinator nodes. The process of routing involves sending broadcast message during path search and establishing a path and sending data along the established path. During the path discovery from the source node to the sink node, it finds a path with maximum amount of residual energy and minimum hop count.

Path with maximum amount of energy is computed by sending the broadcast message to it neighbors. The neighbors send their position and energy information to its source node. The node then establishes the path with the next nieghboring node having highest amount of energy. This process repeats until it reaches the base station. Figure 2a shows the topology with energy information in it. Node 7 is the source node and node 5 the destination. Figure 2b displays the process of broadcasting from node 7. In Figure 2c the path is established upto the node 3 and broadcast message from node 3 to its neighbors. This process is repeated till it reaches the basestation. Figure 2f illustrates the path established up to the base station such that it is along the path with neighbors having high energy. The same is presented for a topology of 25 nodes, with node 0 as the source and node 12 as the sink in Figure 2g. Using the concept explained above in this diagram, the path that is taken to reach the sink, is longer than what is expected.

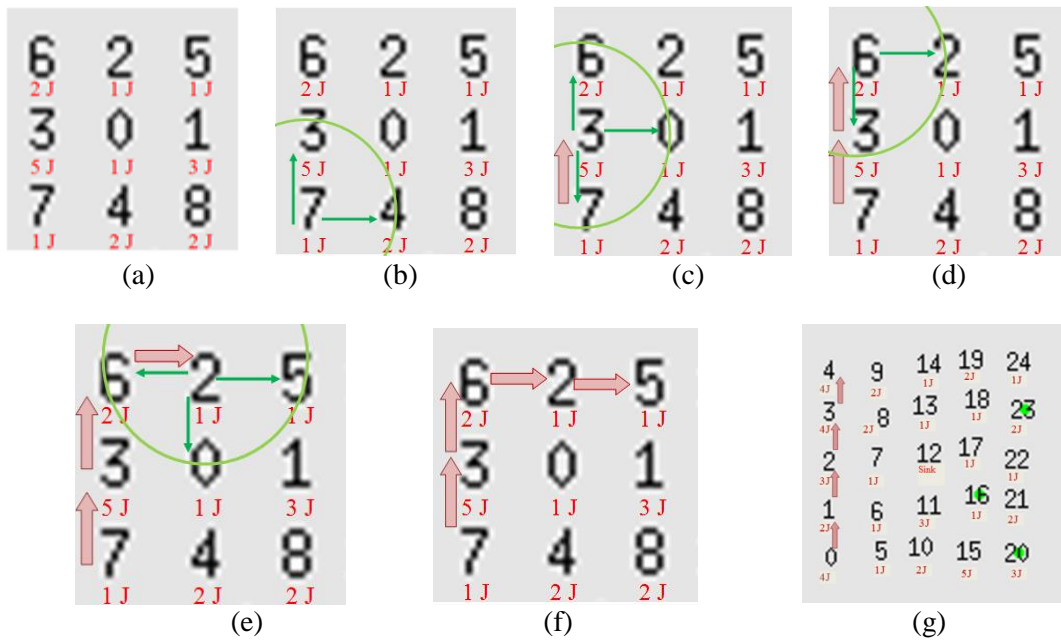

Figure 2 Path establishment with nodes having maximum energy along its neighbors

Alternate strategy is to use the directional information for path finding. Figure 3 shows node 0 is the source and node 12 the sink. [23] address the routing based on directional information. With vertical distance of X from the sink and Y the horizontal distance as depicted in Figure 3.1a the minimum distance to the base station from the source is $\sqrt{(X^2+Y^2)}$

The path is correctly followed if from any node,

(Distance traversed so far from the source node + minimum distance from current traversed node to destination) ≤ (horizontal distance (X) + vertical distance (Y) from the source node to the destination). (1)

At source node 0, using eq. (1) we have,
$(0 + \sqrt{(X^2+Y^2)}) \leq (X+Y)$ (2)

If a path of length $Y_1$ is traversed along vertical direction and $X_1$ along horizontal direction then the distance from the source to the destination is, $X_1+Y_1+$ minimum distance from current node to destination. Using eq. (1) we should have,

$X_1+Y_1+ \sqrt{((X-X_1)^2 +(Y-Y_1)^2)} \leq (X+Y)$ for a valid path (3)

In Figure 3.b, $X_1=0$ and the criteria for valid path is satisfied with,

$Y_1+ \sqrt{(X^2 +(Y-Y_1)^2)} \leq (X+Y)$
In Figure 3.c, with $X_1=0$ and the criteria for valid path is not satisfied as,
$Y_1 + \sqrt{(X^2+Y_2^2)} > (X+Y)$ (4)

Path shown in Figure 3c is longer as it does not satisfy the eq. (1) and those paths are freed during path discovery.

Figure 3.d shows that a path of length $Y_1+X_{11}$ is traversed at node 11. On broadcast from node 11 it reaches node 16. At node 16, from the control packet the knowledge of the distance $Y_1+X_{11}$ so far traversed is obtained. Knowing the node 11's x and y coordinates; the distance $X_{12}$ between 11 and 16 is computed. $X_1 = X_{11}+X_{12}$. The total distance traversed from the source 5 to the current node becomes $X_1+Y_1$. The distance between node 16 and node 12 is computed knowing the coordinates of the current node and that of base station.

The path become invalid as $(X_1+Y_1+ \sqrt{(X_{12}^2 + (Y- Y_1)^2)}) > (X + Y)$.

Paths with highest amount of residual energy along its route could result in a long lasting path. Given the energy levels $e_{i1}, e_{i2}, e_{i3}\ldots e_{in}$ in each of the 1..n nodes along the path i, where 1st node represents the source and nth node the destination node

the amount of total residual energy along path i is given by $E_i = \sum_{k=1}^{n} eik$ (5)

If another path has energy levels $e_{j1}, e_{j2}, e_{j3}\ldots e_{jn}$ in each of the nodes 1..n with 1 the source node and n the destination along the path j, the amount of residual energy along path j is given by $E_j =$

$\sum_{k=1}^{n} ejk$ (6)

Path is so chosen such that, for all the paths p from 1...m satisfying the minimum hop count, using eq. (5) and eq. (6), the path with maximum residual energy is given by,
Max. {Ep} $\forall$ path p from 1…m. (7)
where Ep represents total residual energy along path p.

Figure 4a illustrates the topology with source node 7 and destination node 5, with residual energy shown against each node. Path shown in Figure 4b is with neighbour having maximum energy. Figure 4c is path with maximum total residual energy. (1+5+1+3+1=11J). The best path is one with higher minimum energy along its path as shown in Figure 4d.

The minimum energy node is the first one to lose its energy. Hence the higher the amount of minimum energy along the path, the longer the period the data gets transmitted. Hence a third approach is used which has maximum amount of minimal energy along its path.

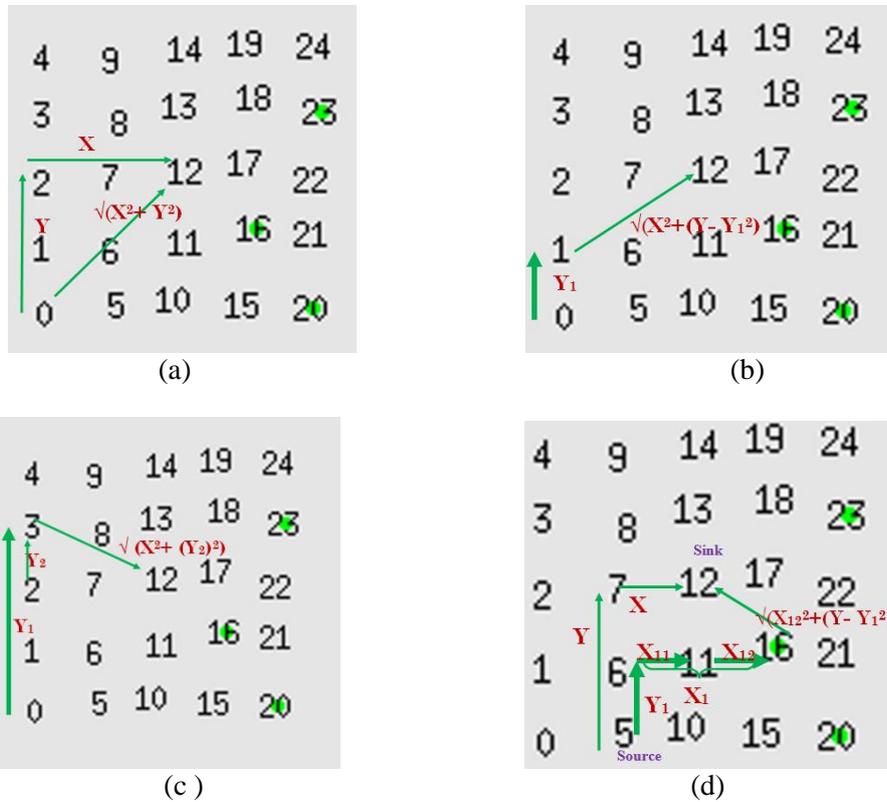

Figure3: Various Partial Paths Traversed (a) Optimal Path (b) Path with Minimum Hop Count (c) Path with Higher Hop Count (d) Path with a Different Source and Higher Hop Count

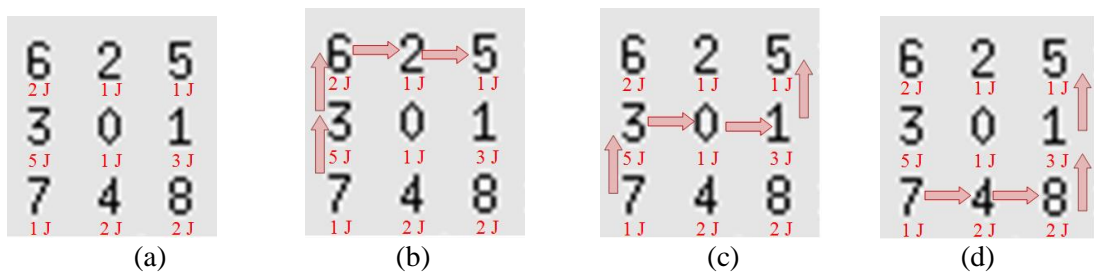

Fig. 4 Routing methods

A path containing the maximum amount of minimum value of energy in the node is chosen as the best path. Longer life of a path is based on the first node losing its energy which is the node with minimum energy. Hence chose a path with higher value of minimum energy.

Given the residual energy levels of $e_{i1}, e_{i2}, e_{i3}\ldots e_{in}$ in each of the 1..n nodes along the path i, where 1st node represents the source and nth node the destination node, the minimum residual energy level along the path i is given by

$E_{mini} = \min\{ e_{i1}, e_{i2}, e_{i3}\ldots e_{in} \}$ (8)

If another path has energy levels $e_{j1}, e_{j2}, e_{j3} \ldots e_{jn}$ in each of the nodes 1..n with 1 the source node and n the destination along the path j, the minimum residual energy level along the path j is given by

$E_{minj} = \min\{ e_{j1}, e_{j2}, e_{j3} \ldots e_{jn} \}$ (9)

The best path is the one with min hop counts and highest value among the minimum energy value. If m path exists then the path chosen is such that it satisfies max{ $E_{min1}, E_{min2}, \ldots E_{minm}$ }

(10)

If there exists many paths with same maximal minimum energy, then path with higher residual total energy along the path is considered. Figure 5 shows the three paths to send the data from node 0 to node 24. The path along 0→5→10→15→20→21→22→23→24 is the maximum total residual energy path with value (4+1+1+1+8+1+1+3+2=22J). The path along 0→1→2→7→8→13→18→19→24 and 0→1→2→7→8→13→18→23→24 are paths with maximum- minimal residual energy among the various path from source to destination. Both the path has minimum residual energy node of value 2J, but the path passing through the node 23, has higher total residual energy among them and this path is chosen. Path containing less than the minimum threshold value to sustain at least one data packet dispatch is dropped.

An algorithm is proposed to carry out maximum amount of minimum energy. Every node has a routing table as shown in the Table 2. This routing table helps in establishing a path from any source node to the destination. To achieve this control signals of the type in Figure 6 are broadcast from the source node to its neighbors. The working of the basic protocol is explained in my earlier paper [21]. The packet Type PATH_SEARCH. Each node in the routing table maintains the maximum value of minimum energy of any path that has traversed through that node.

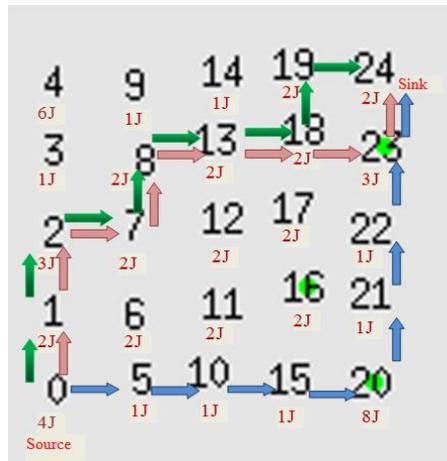

Figure 5 Paths traced by various protocols

```
 0                   1                   2                   3
 0 1 2 3 4 5 6 7 8 9 0 1 2 3 4 5 6 7 8 9 0 1 2 3 4 5 6 7 8 9 0 1
+-+-+-+-+-+-+-+-+-+-+-+-+-+-+-+-+-+-+-+-+-+-+-+-+-+-+-+-+-+-+-+-+
|Packet Type  | Sequence no |  Hop Count  |                    |
+-+-+-+-+-+-+-+-+-+-+-+-+-+-+-+-+-+-+-+-+-+-+-+-+-+-+-+-+-+-+-+-+
|                       Source Address                          |
+-+-+-+-+-+-+-+-+-+-+-+-+-+-+-+-+-+-+-+-+-+-+-+-+-+-+-+-+-+-+-+-+
|                     Destination Address                       |
+-+-+-+-+-+-+-+-+-+-+-+-+-+-+-+-+-+-+-+-+-+-+-+-+-+-+-+-+-+-+-+-+
|                        Minimum Energy                         |
+-+-+-+-+-+-+-+-+-+-+-+-+-+-+-+-+-+-+-+-+-+-+-+-+-+-+-+-+-+-+-+-+
```

Figure 6. Path discovery with Minimum Energy Information

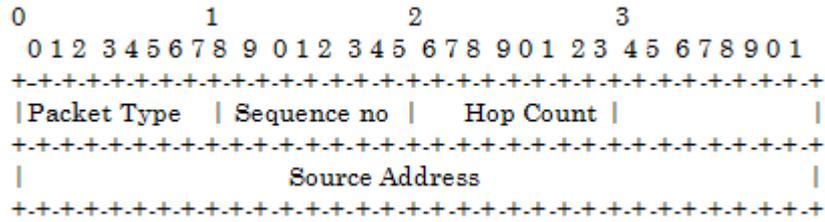

Figure 7 Packet format used for Path Establishment

Table 2: Routing Table

| Information stored | Size occupied in bytes | Functionality |
|---|---|---|
| Destination Address | 4 | For path discovery the forward route destination address |
| Next Hop Node Address | 4 | Next neighbour address towards destination |
| Previous Hop Node Address | 4 | Previous neighbour address towards source |
| Time Stamp | 4 | Time when the signal was received |
| Sequence Number | 1 | The number of times path search was carried out |
| Hop Count | 1 | Hop count from the source |
| Route_establish_flag | 1 | Set to zero during path discovery and set to one on path establishment. |
| Address of Neighbours* | 4 | Address of the linked list containing the information of its neighbours |
| Source Address | 4 | Address of the node for which path discovery is initiated |
| Minimum Energy | 4 | Minimum Energy along the path (not including the current node energy level.) |

## 5.1. Routing Algorithm for maximum amount of minimum residual energy

a) Start the path search using the control signal with a new sequence number from the source to the destination using the control signal of the type in Figure 6.
b) In each of the traversed node
   i) Drop the control packet when it encounters a node which has energy less than the path threshold energy level (Energy to transmit control signal and at least one data signal).
   ii) Make an entry in the routing table for each source destination combination for each of the node traversed provided,
      i. The route to be searched is a new one with a new sequence number
      ii. With minimum hop count from those found so far between the corresponding source destination combination
      iii. Distance traversed so far from the source node + minimum distance from current traversed node to destination < horizontal distance+ vertical distance from the source node to the destination
      iv. Having maximum minimal energy from a path traversed through the same node.
      v. Having maximum total energy.
      vi. For each entry made in the routing table a reverse routing table entry is made from current traversed node to its previous source node.
      vii. Otherwise drop the control packet.

c) If the hop count is more than the maximal hop count that can be supported in this network, drop the packet as this gives an indication that there is no route from this node to the destination.
d) This process is repeated until it reaches the destination to drop the control packet.
e) Once the packet reaches the destination another control packet of the type in Figure 7 is started in the reverse direction from the destination to the source.
f) The node traverses the path as specified in the reverse routing table to reach the destination.
g) Once the control signal reaches the source, data transmission starts from the source to the destination in the path as established during the reverse path establish.

***Inference of algorithm 5.1:****Path having maximum amount of energy along its path can sustain for long intervals of time and avoids repeated transmission and reception during path search. In this algorithm, the path containing maximum amount of minimal residual energy along with maximum total residual energy and directinal information is considered for establishing the path.This path once established will run for longer intervals of time.*

## 6. AGGREGATION

Certain area of the agricultural land may not be even. This could result in deploying more sensors around those area. In such cases, it may have related value with the neighboring sensors. If all the associated data is sent to the base station at the same time large amount of energy could be consumed because of the repeated transmission of correlated data. This results in draining of the energy in the battery. To avoid this, in-network aggregation is carried out. Various aggregation techniques are carried out. Data aggregation along the spokes of the wheel is proposed[24].

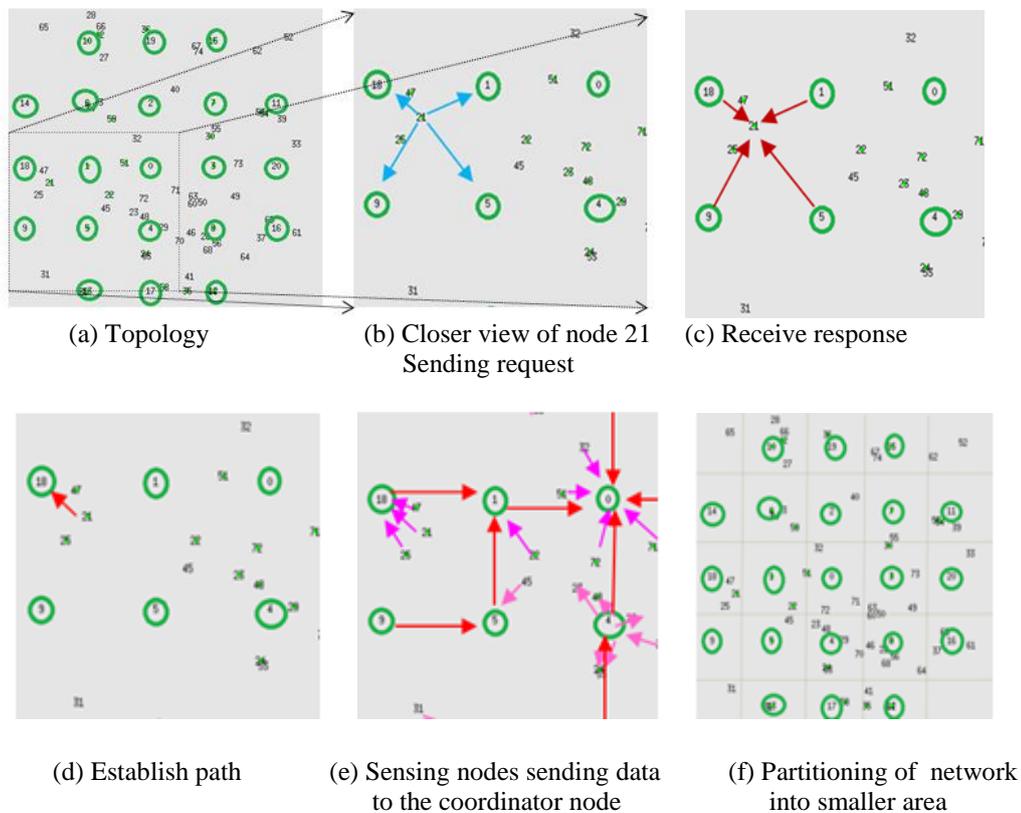

(a) Topology  (b) Closer view of node 21 Sending request  (c) Receive response

(d) Establish path  (e) Sensing nodes sending data to the coordinator node  (f) Partitioning of network into smaller area

Figure8. Various stages to establish a path between the sensing and its coordinator node.

155

In the current study, aggregation is carried out to find a single value like the minimum or the maximum in the whole network. It is also used to find an average value in a particular area of the network.

Figure 8 depicts the various stages for searching the coordinator node. Nodes circled with green colour represent the coordinator nodes. These are placed as per the deployment strategies of Figure 1. Other nodes are non-coordinator nodes which are sensing and sending their values to the coordinator nodes. The coordinator node can sense, receive, compute and forward the data to the next coordinator neighboring node.

Figure8a, picturises the deployment of the sensor nodes. Figure 8b, the close up of a small area with one particular node with id 21 sending the sensing request information to the nearby coordinating nodes. Node 0 is the basestation. Node 18 and node 1 are along the path established to the base station.

The packet details are as per Figure 9a. Figure 8c shows the sensing reply packet received from the four neighboring coordinator nodes to the node id 25. Reply packet format is as shown in Figure 9b. Node id 25 finds the coordinator node based on the least distance towards the coordinator node along with the maximum residual energy among the coordinator nodes and sends back the message to one of the coordinating node as shown in Figure8d. The non coordinator node's information is updated in the coordinator node. This process is carried out along the whole network. Figure 8e shows the non-coordinator nodes sensing data and aggregating its value with its coordinator nodes along the path 18→1→0. Figure8f shows the partitioning of the network into sub areas with one coordinator node in each.

In the packet format of Figure 9, Packet type, is CORD_SEARCH packet type. Sequence no is incremented, each time it is searching for a new coordinator. Source Address is the address of the node trying to search for coordinator. In Figure10 the source address corresponds to the coordinator node, with Coordinate Node x and y position specifiying the position of the coordinator node. This information is sent to the non-corrodinaotr node from the coordinator node. This packet size is smaller than that used to search for the next hop neighbor as in AODV reducing the amount of energy consumed.

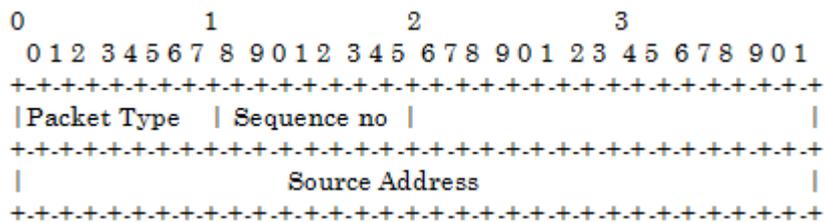

Figure 9 Packet Format to Search for the Coordinator Node

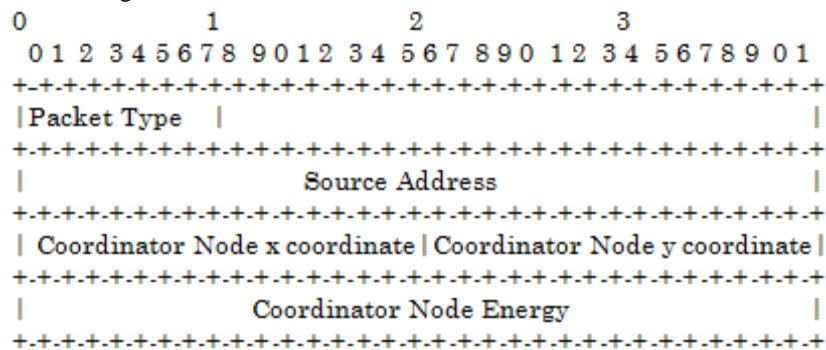

Figure 10 Packet format to establish the path.

Once the path to the coordinator is found, the next step is to accept data and send it to the base station.

If all the nodes in the network send their data to the base station, there is likelihood of large amount of energy consumption. In order to avoid this, two types of aggregation are discussed. In the first type, the minimum data (soil moisture content) information in the whole of network is sent to the base station as shown in Figure 11a and explained with the algorithm in section 6.1. In the second case, Figure 11b the aggregated averaged data value of a particular area is sent to the base station and explained with an algorithm in section 6.2

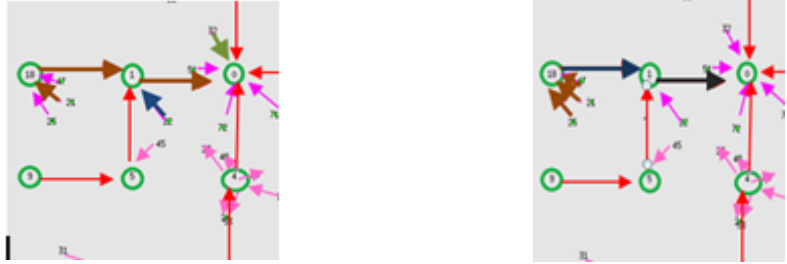

(a) Sending the minimum data of the whole network   (b) Average value around around a coordinator node

Figure 11 Routing for sending data values to the base station

## 6.1. Minimum data value in the whole network

To find minimum data in the whole network, consider the Figure 12 with coordinator node $C_1$, with non coordinator nodes as $S_{11}$, $S_{12}$, $S_{13}$, with data values $d_{11}$, $d_{12}$, $d_{13}$. The minimum data among these, $dmin_1 = \min\{d_{11}, d_{12}, d_{13}\}$ is updated in the routing table for the node $C_1$ provided, the data is the current data. Information pertaining to minimum data value node id and the location information is updated in the routing table. Node $C_1$ on sensing data $d_1$ is compared with $dmin_1$. On updating the routing table with $dmin_{12} = \min\{dmin_1, d_1\}$ in node $C_1$, send the data along with the position information of minimum data value node to the next hop node $C_2$. Node $C_2$ updates its routing table to the value $dmin_2=\min\{d_{21}, d_{22}\}$. $\min\{dmin_2, dmin_{12}\}$ is updated in the routing table. Node $C_2$ on sensing data $d_2$ updates its routing table with the value which is the minimum of all data at that node, i.e. $\min\{\min\{dmin_2, dmin_{12}\}, d_2\}$ and forwards this data to the next neighboring node. This process is repeated until the messages reaches the base station, sending along with it, the information of the node id with minimum data value, its time of sensing the data and the position of the node.

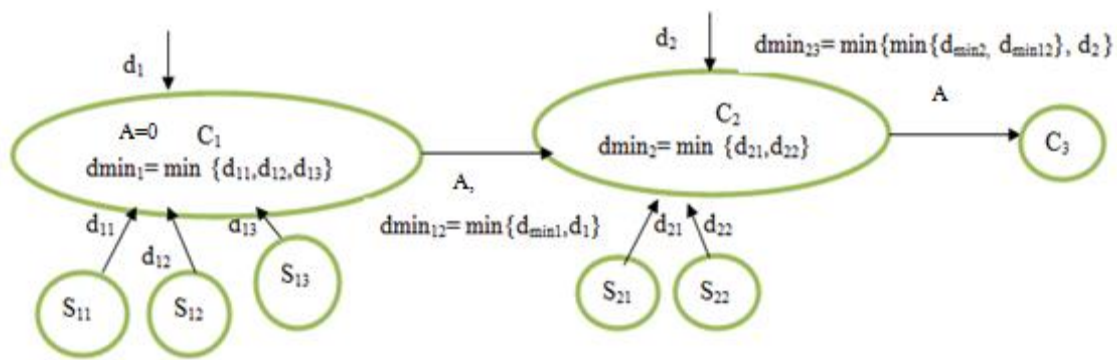

Figure 12 Data aggregation at the coordinator node which is sent to base stations aggregating along its path.

*Algorithm 6.1*
   a) Non coordinator node routing table is updated with the values obtained from the application layer. These values include data from the sensing node, position of the node, the node id, time when data is sensed and energy associated with the node.

b) The non–coordinator node data on sending data to the coordinator node, the neighboring list information is updated in the coordinator nodes, provided it is the recent data and this information packet is dropped.
c) Get the information of all the neighboring non –coordinator nodes and compare it with the coordinating node and update in the coordinator nodes routing table with the value of least data among all these.
d) When data is forwarded to the next coordinator from the previous coordinator node, update the previous coordinator nodes information in the current node.
e) Get the node with the minimum data information among its neighbors, (if neighbor exists) with current information, assign it to the current nodes routing table.
f) If the current node, senses new data and if this data is less than the routing data, update the routing data with the current sensed data.
g) If the data sensed is larger than the routing data, with routing table having current data information, update the sensed data information with the routing table data information.
h) The data packet obtained from the previous node is dropped.
i) The process of comparing with the neighboring nodes and forwarding data is carried out until it reaches the base station.
j) At the base station the aggregated information of the whole network i.e., the minimum data value with its node id, position, energy and the time when the data was sensed along with the aggregation type (0) is available.

*Inference of algorithm 6.1:* Aggregation results in sending minimal information to the base station. This is due to the fact that large amount of data is partially computed in the coordinator nodes. The minimum data in the whole network helps the operator of the network to regulate the water supply to that particular area of the network.

## 6.2 Aggregated value at any specified coordinator node (or specific area of the field)

To find average data value in a particular area at the location of the the coordinator node $C_1$ in the network, the average value $davg_1 = avg\{d_{11}, d_{12}, d_{13}, d_1\}$ is computed and sent to the next hop node. This data is forwarded without updating the routing table information along the route towards the base station. This procedure is depicted in the Figure 13.

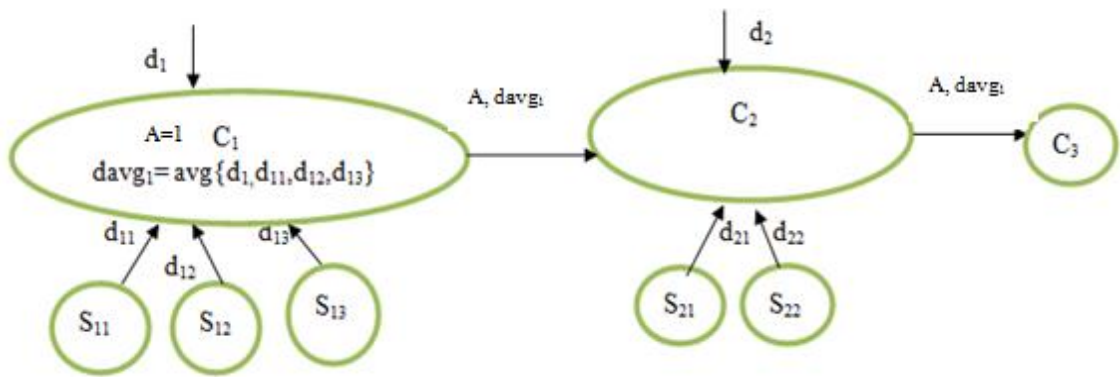

Figure 13 Computation of average value of the data at the location where node $C_1$ exists.

*Algorithm 6.2*

a) Non coordinator nodes obtain the sensed data informatin and updates those values in their routing table. These values include data from the sensing node, position of the node, the node id, time when data is sensed and energy associated with the node.

b) When the data reaches the coordinator node, recent data obtained from the non-coordinator node is updated in the neighboring node information list and the packets are dropped.
c) The coordinator node averages the data from its entire non-coordinating node along with its information and sends it to the next hop neighbor.
d) The process of forwarding is done till it reaches the base station where the information of the coordinator node, with the averaged data, the time of aggregation, position of the coordinator node and the aggregation type (1) is obtained.

*Inference of algorithm 6.2:* This algorithm gives the average soil moisture content at certain area of the network. This helps in analysing the water supply in different parts of the field. Water distribution as used with sprinkler or any other method like using pump, which had to be computed at the base station is partially carried out within the network.

## 6.3 Triggering of event in the case of failing battery supply or water level going below | above the required threshold

Failing battery sends a notification to the the base station. If the battery level goes below the threshold level ethresh, then a warning is sent to the base station carrying with it the information pertaining to the energy level and the location of the sensor. This could assist the manager to change the battery if it is a crucial node. If $e_{11}$, $e_{12}$, $e_{13}$ are the energy values of the child nodes and $e_{10}$, that of the coordinator node, any node whose energy level is below the threshold will report to the next coordinator node and the process is repeated until it reaches the base station. The node whose energy level is below threshold along with its position is obtained at the base station. This process is shown in the Figure 14. The same concept when used with the water level indicator is shown in Figure 15.

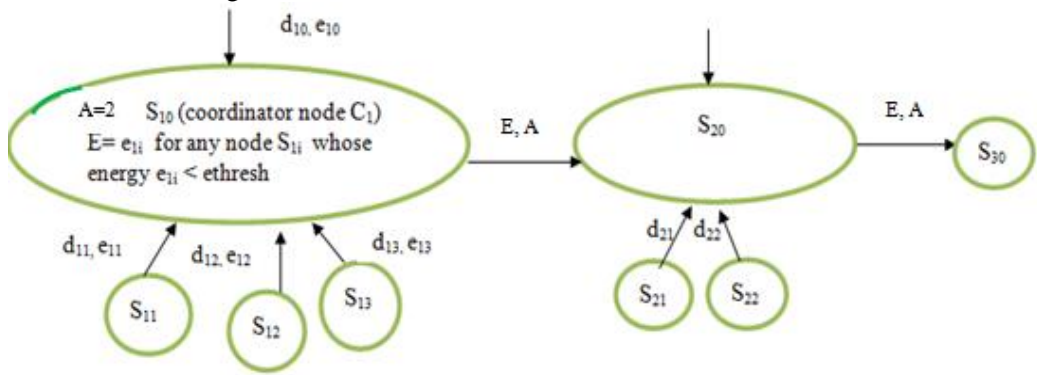

Figure 14 Reporting the energy drained level to the base station

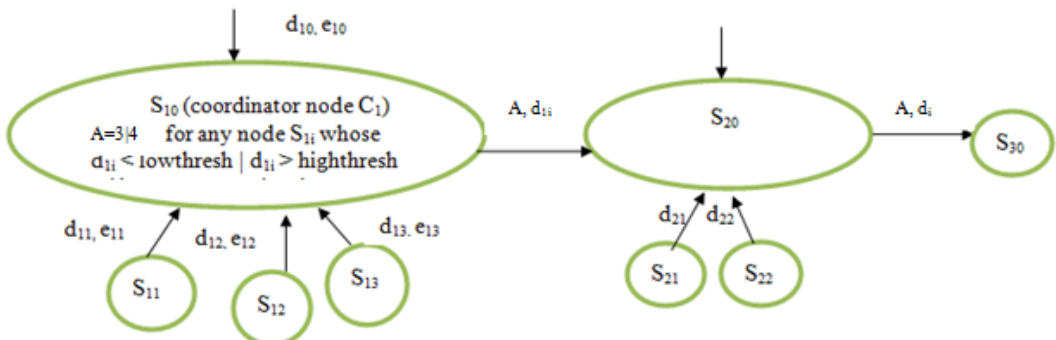

Figure 15 Water threshold indicator

*Algorithm 6.3*
   a) Each non coordinator node node provides information of the data sensed along with its energy information to the coordinator node.
   b) If the non-coordinator data/energy is below the threshold, information is  passed to the coordinator node and the packet is dropped.
   c) An event is triggered if the non-coordinaotr node or coordinator node is below/above the threshold either for the energy or water.
   d) If the path is not established to the base station, a new path is established to the base station from the coordinator node and data is  immediately sent without waiting for the nodes along the path to sense their data.
   e) Triggered information type whether it is battery(3) or water(4|5), remaining energy value, data value, position and the interval when the reading was taken is intimated to the base station.

*Inference for algorithm 6.3:*
   a) **Battery Depletion:** *This algorithm generates an event when the energy level goes below the threshold level of handling even a single data transmission. If they are non-coordinator  sensing nodes, an immediate message is passed on to the base station without waiting  for the coordinator node to initiate the transmission.  If it is a coordinator node the battery low message is passed on as soon as the data is sensed with higher priority. This could give an indication to the operator at the base station, the location of the node whose battery is almost drainied. This gives an opportunity for the operator to change the battery if the location of sensing is crucial or the self-organising network will find an alternate path.*
   b) **Water Level requirement:** *All nodes need not send message to the base station on timely basis. The requirement of notifying the water level is when it is in excess or shortage of water. This procedure can put all the nodes to sleep accept the ones responding to events. The disadvantage being such events occuring rarely could make the operator unaware if the network is dead or alive.*

# 7. SIMULATION RESULTS

The simulation is carried out using ns2.34.  Simulation parameters utilized in this work are as per the Table 3 complying with Table 1.  The comparison with minimum hop with total residual energy is carried out in the earlier work [22]. The designing of the parameters for simulation is as generated in the Table 3 is also addressed in this reference paper. In the current work, the time interval for transmission of data is once in every 300 seconds. The simulation is executed for a period of 6900secs. The source is node 0 and destination node 24 for topology as in Figure 5.

Table 4 shows the comparison of two proposed routing protocols.  In the first protocol the maximum amount of minimum residual energy along with total residual energy along with the directional information is used for computing the route. In the second proposed protocol only the maximum total residual energy is taken into account. It is observed that though many transmissions take place during path establishment in the first case, the path established by this method is alive for a longer interval of time with less amount of energy consumed. The extra energy consumed in the second protocol was to find if any other path exists to reach the destination. In this set up there is not much of a difference in the energy level between various node energy level, hence only one extra transmission has taken place, else network could run longer for the established path.

Table 3 Simulation parameters

| Radio Parameters | | Simulation parameters | |
|---|---|---|---|
| Radio frequency | 868 MHz | Date acquisition interval | Varying intervals of 150/300/600seconds. |
| Antenna Height | 1m (min. reqd. height 0.0819m) | Nodes | 21 coordinator nodes, 54 non- coordinator nodes |
| Antenna Type | Omnidirectional – Quarter wave | Topology | Square/Hexagonal |
| Transmit Power | 3.16 mW =5dBm | MAC | 802.11 |
| Receive Power | −104dBm@ 5dBm=3.98e-14W | Queue | Drop Tail |
| Carrier Sense Threshold | −104dBm@ 5dBm | Queue size | 50 |
| Capture Threshold | 10 dB | Protocol | Proposed aggregation algorithms |
| Gain of transmitting and receiving antenna | 1 | Transmission range | 528m |
| | | Simulation period | 1067110 seconds nearly 13 hrs.used for aggregation protocol |
| **Sensor parameters (Tiny Node)** | | **Battery (Alkaline battery of 1.5V)** | |
| Transmit Power | 0.099W=19.95dBm | Battery supply | 3V with 2 AA sized alkaline battery |
| Receive Power | 0.042W=16.23dBm | Power consumption | 0.0705W for 23.5mA discharge current |
| Sleep Power | 0.000003W= -25.2 dBm | Energy consumption | 20304J for 80 hrs. of active operation |
| Idle Power | 0.006W =7.78dBm | | |

Table 4: Comparison of routing protocol for consumption of energy and delivery ratio

| **Square Topology with 25 nodes.** | | **With Max-min energy+ maximum total residual energy + directional information** | | **With maximum total residual energy** | |
|---|---|---|---|---|---|
| Case I 24 data packet sent | Packet delivery ratio | 23/24 | 95.8% | 22/24 | 91.6% |
| | Total average energy consumed (J) | 7.9242 | 7.92043 –I$^!$ 0.00206 –T$^*$ 0.00178 –R$^\#$ | 12.17217 | 12.16840–I 0.001912–T 0.001853– R |
| Case II One data packet sent | Packet delivery ratio | 1/1 | 100% | 1/1 | 100% |
| | Total average energy consumed (J) | 0.00204 | 0.00095 –I 0.00050 –T 0.00059 –R | 0.001707 | 0.001027–I 0.000312 –T 0.000367 – R |

All energy measurements are in Joules.
Split up of total energy is shown as  I$^!$-Idle energy T$^*$-Transmit energy R$^\#$-Receive energy

**Inference from the result:***The path search process consumes more energy using the maximum amount of minimum residual energy along with directional information and total energy residual path than using only the total maximum residual path. But path once established it runs longer as the path has higher minimal energy level. This will make the first node along the path to die at a much later time than with just total maximum residual energy.*

To carry out aggregation coordinator nodes are deployed using grid topology. Other non-coordinator nodes are dispersed randomly. Figure 16 shows the topology of the two types of deployment used.

Ns2 does not support data handling directly. In order to incorporate handling of data, the simulator is extended both in the application layer and the UDP layer. Providing data value for all the 75 nodes is cumbersome, so inputs are stored in the files. Each file has information pertaining each of the individual sensors, the data sensed by the sensor, along with the type of aggregation required (average=1 or whole network=0) and time instance relative to current time when the next type of aggregation has to be carried out.

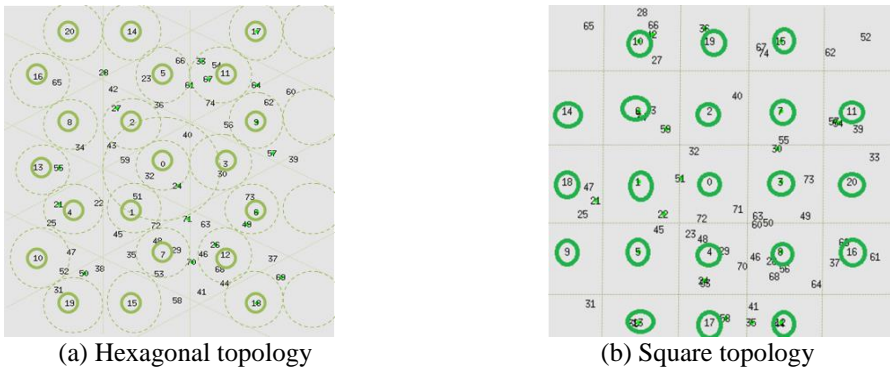

(a) Hexagonal topology    (b) Square topology

Figure 16 Topology showing the region for aggregation in the block area.

If the input data is changing rapidly as in the case of rainfall or during irrigation the rate at which data is read is fast (every 2.5 minutes=150s). If there is no water supply to the field the readings are taken once in 10 minutes (600s). The network does not read data at night times. All the nodes sense and send data to their neighbour node. It is multisource single destination concept. During path search if a path is already established, nodes along the path when they sense the data, utilize the existing path, instead of finding a new one.

A sample input reading of one of the sensor is shown in Table 5 and the output information of the whole network stored at the base station is shown in Table 6.

Table 5 Example of the input file for one of the nodes –node id 9

| Data value in cm | Aggregation type (1 for average 0 for whole network) | Relative time interval in seconds |
|---|---|---|
| 2.90 | 0 | 0 |
| 2.92 | 0 | 600 |
| 2.93 | 0 | 600 |
| 2.95 | 0 | 600 |
| 2.104 | 1 | 300 |
| 2.208 | 0 | 300 |

Table 6 Sample output at the base station

| Time at base station in seconds | Time when data was read in seconds | Id of the node sending the data | Sensor data reading in cm. | Position of X coordinate in meters | Position of Y coordinate in meters | Agg. Type* | Energy in Joules |
|---|---|---|---|---|---|---|---|
| 2.06204 | 1.002000 | 51 | 1.060 | 842.0 | 1080.0 | 0 | 0.281649 |
| 602.061 | 601.00200 | 21 | 1.800 | 175.0 | 1000.0 | 0 | 20303.980 |
| 1202.06 | 1202.0000 | 18 | 2.010 | 0.0 | 1056.0 | 1 | 20303.975 |
| 1802.06 | 1802.0600 | 1 | 2.004 | 528.0 | 1056.0 | 0 | 20303.953 |
| 2102.06 | 2102.0600 | 9 | 2.104 | 0 | 528.0 | 1 | 20303.980 |
| 2402.06 | 24001.002 | 51 | 2.000 | 842.0 | 1080.0 | 2 | 0.025904 |

Aggregation type*  0-Average, 1- Whole network,  2- Low battery, 3- Low water level, 4- High water level

The data received at the base station is stored in a file, the information conveyed to the base station is the type of aggregation that was carried out (average or whole network) or the battery discharge information or water access or shortage triggered information along with the time the aggregation /event occurred and the position of the node which is conveying the information.
In Table 7 in column I, all the data are sent via the coordinator node, and this has higher amount of energy consumption (11.7998J). In column II, the data is aggregated at each of the coordinator node and packets are dropped after it sends the information to the next node and hence the amount of energy consumed is the minimum (6.743J). The last column displays the amount of energy consumed in associating the non-coordinator node to the coordinator node with energy consumption of 0.01454J. The first 2 columns are run for a time period of nearly 13 hours (1067110 seconds). The last column is for 30seconds when the non-coordinator node associates with the coordinator node.

**Inference as obtained from the results:** *It is observed that dropping the packets after it is sent to the next hop neighbor or coordinator node saves considerable amount of data, instead of all the nodes sending data to the base station using their own path*

Table 7 Relative comparison of consumption of energy at various levels of aggregation

| All readings are in Joules | Column I | | Column II | | Column III | |
|---|---|---|---|---|---|---|
| | Finding coordinator node and transmitting all data | | Finding coordinator node, aggregating and selectively transmitting | | Finding coordinator | |
| Case I Square topology | 11.7998 (Total energy) | 0.0000 – I 1.5902 – T 10.209 – R | 6.743 (Total energy) | 0.0000– I 0.9644– T 5.7788– R | 0.01454 (Total energy) | 0.00000 – I 0.00207 – T 0.01247 – R |
| Case II Hexagonal topology | 11.0564 (Total energy) | 0.0000 – I 1.7376 – T 9.3188 – R | 6.409 (Total energy) | 0.0000– I 0.9638– T 5.4459– R | 0.013159 (Total energy) | 0.00000– I 0.00213 – T 0.01102 – R |

## 7. DISCUSSION AND CONCLUSION

Sensors are deployed in the agricultural field to monitor the moisture, humidity and temperature. It is essential that the batteries have sufficient energy to transmit data for the entire cropping period. To accomplish this, efficient routing and aggregation techniques are essential so as to extend the life of the battery.

In this paper, different ways of establishing the path is studied. The path passing through nodes should be such that that they have maximum amount of energy among all the different paths reaching the base station from the source node. The path should be such that the path once established is sustained for long intervals of time. The advantage of having such an approach is that, the number of times the new path is searched gets reduced. This results in saving energy which would have been expended because of unnecessary transmission and reception of control signals. One of the approaches is to find the neighbour with maximum residual energy. The next approach is the path which has the maximum total residual energy. Other approach is to follow the path with maximum amount of minimum residual energy along the path. Directional information also reduces a lot of unnecessary broadcasting of the control signals. All these approaches are used to find efficient path from the source to the destination.

Aggregation techniques are used to avoid transmission of related data by aggregating them at the coordinator node before transmission. This saves the processing time and computation at the base station. Two types of aggregation are discussed; one is to find minimum data across the whole network. Other is to find average value along a certain area of the network. Events are also triggered when the battery level goes down or when the water supply exceeds the required threshold or below desired value. The operator at the base station can take necessary steps to handle the situation. Non-coordinator nodes are associated with the coordinator nodes based on the minimum distance and maximum residual energy among the neighbouring coordinator node. Higher the energy at the coordinator node, more number of nodes are associated with the coordinator node. This balances the network which has uneven distribution of energy.

This work can be enhanced in the future by aggregating the data values based on closely related information rather than the distance and energy information. This paper emphasizes the need of finding efficient routing and aggregation techniques for the network to live long without wasting energy on unnecessary transmission and reception. Having a proper decision support system could help the farmers to take necessary action to monitor the way water is managed and enhance the yield.

**Authors**

Smitha N. Pai is an Assistant Professor in theDepartment of CSE at MIT Manipal. She obtained her M.Tech. in CSE from ManipalUniversity in 2002 and B.E in E&C fromMysore University. She is currently pursuing her PhD. from Manipal University. Her main interest lies in wireless sensor, adhoc networks.

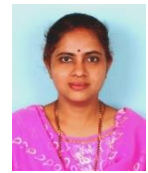

**Dr.K.Chandrashekar Shet** obtained his B.E, M.Sc. (Engg) andPh.D degrees from Mysore University, Sambalpur Universityand IIT Bombay in the years 1972, 1979 and 1987 respectively.He is working in NITK Surthakal since 1980 and presently he is the professor in the Dept. Of Computer Sc. & Engg, and Dean(Faculty welfare).He has published around 250 papers, in National, International,journals/conferences. Besides, he has published three books, on Micro-processors, Software Engg. & quality Assurance. He has produced 10 Ph.D professionals and currently six are pursuing research under his guidance leading to Ph.D.

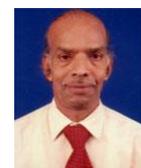

**Dr. H. S. Mruthyunjaya** has completed his bachelor degree in Electronics and Communication Engineering from Mysore University in 1988 and obtained his masters degree in Electronics and Control Systems Engineering from Birla Institute of Technology and Science, Pilani in 1994. He has a Ph.D in Electronics and Communication Engineering conferred by Manipal University for his thesis entitled 'Performance Enhancement of Optical Communication Systems and Networks using 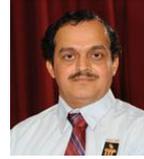
Error Control Techniques'. He is currently serving as a Professor in the Department of Electronics and Communication Engineering, Manipal Institute of Technology, Manipal, India where he joined as a Lecturer in the year 1998. He has done research on countering non-linear effects and other noises in WDM all-optical networks by employing error control coding techniques. His areas of major interests are the Optical Fiber Communication systems, Fiber Optics, Photonic Crystal Fibers, WDM networks and systems, Electromagnetic theory & General areas of Digital Communication Systems. He has authored or co-authored over Forty three technical papers in refereed journals and International conference proceedings. He is a Fellow of the Institution of Engineers (India) and member of Indian Society for Technical Education.